\documentclass[10pt,letterpaper]{article}
\usepackage{opex3}

\begin{document}

\title{An optically trapped mirror\\ for reaching the standard quantum limit}

\author{Nobuyuki Matsumoto,$^\ast$ Yuta Michimura, Yoichi Aso, \\and Kimio Tsubono}

\address{{\it Department of Physics, University of Tokyo, 7-3-1 Hongo, Bunkyo-ku, Tokyo 113-0033, Japan}}

\email{$^\ast$matsumoto@granite.phys.s.u-tokyo.ac.jp} 


\begin{abstract}
The preparation of a mechanical oscillator driven by quantum back-action is a fundamental requirement to reach the standard quantum limit (SQL) for force measurement, in optomechanical systems.
However, thermal fluctuating force generally dominates a disturbance on the oscillator. 
In the macroscopic scale, an optical linear cavity including a suspended mirror has been used for the weak force measurement, such as gravitational-wave detectors. 
This configuration has the advantages of reducing the dissipation of the pendulum (i.e., suspension thermal noise) due to a gravitational dilution by using a thin wire, and of increasing the circulating laser power. 
However, the use of the thin wire is weak for an optical torsional anti-spring effect in the cavity, due to the low mechanical restoring force of the wire. 
Thus, there is the trade-off between the stability of the system and the sensitivity. 
Here, we describe using a triangular optical cavity to overcome this limitation for reaching the SQL. 
The triangular cavity can provide a sensitive and stable system, because it can optically trap the mirror's motion of the yaw, through an optical positive torsional spring effect. 
To show this, we demonstrate a measurement of the torsional spring effect caused by radiation pressure forces. 
\end{abstract}



\section{Introduction}
In recent years, significant improvements in optical and mechanical elements have led to the development of the field of optomechanics, where mechanical oscillators couple optical fields via the radiation pressure of light \cite{markus2013}.
As for force measurements such as gravitational-wave (GW) detectors \cite{harry2010,somiya2012}, in which the mechanical oscillator is used as a probe of external force, its sensitivity has almost been limited by the standard quantum limit (SQL), whose (double-sided) power spectrum is given by \cite{khalili2012}
\begin{eqnarray}
S_{\rm FF,SQL}^{(2)}(\omega)=\hbar|\chi_{\rm m}(\omega)|^{-1}+2\hbar\omega_{\rm m} \gamma_{\rm m}m.
\label{sql}
\end{eqnarray}
Here, $\omega$ is the angular frequency, $\hbar$ the reduced Planck constant, $m$ the mass of the oscillator, $\gamma_{\rm m}$ the amplitude mechanical decay rate (i.e., the mechanical quality factor $Q_{\rm m}$ is given by $Q_{\rm m}=\omega_{\rm m}/2\gamma_{\rm m}$) and $\chi_{\rm m}$ the mechanical susceptibility.
Also, theoretical analysis has proven that there is a connection between reaching the SQL imposed on
the free mass (so-called free-mas SQL) and the generation of entanglement states, even between massive mechanical oscillators such as suspended mirrors \cite{muller2008}. 
Because of the massiveness, such states might have a key to investigate effects concerning macroscopic quantum mechanics, such as the gravity decoherence \cite{miao2010}. 
Therefore, optomechanics is not only useful as sensitive probes for the weak force, but also leads to possibilities of testing fundamental problems. 

One of the key milestones toward the SQL is the observation of quantum back-action, whose power spectrum is given by
\begin{eqnarray}
S^{(2)}_{\rm FF,q}&=2\hbar^2\kappa |G_{\rm opt}|^2|\chi_{\rm c}(-\omega)|^2.
\label{qba}
\end{eqnarray}
Here,  $\kappa$ is the total decay rate of the cavity, 
$\chi_{\rm c}$ the cavity susceptibility, and
$G_{\rm opt}$ the light-enhanced optomechanical coupling constant. 
Also, we naturally assume that the linewidth of the cavity is sufficiently larger than sum of the sideband frequency and the cavity detuning. 
The quantum back-action is a measurement-disturbance derived from the Heisenberg uncertainty principle (HUP) \cite{heisenberg1927}, and thus this is the force noise. 
In order to reach the SQL, the force noise should be dominated by the quantum back-action; however, measurements are usually dominated by a strong thermal fluctuating force, whose power spectrum is given by
\begin{eqnarray}
S^{(2)}_{\rm FF,th}=4k_{\rm B}T\gamma_{\rm m}m.
\label{thermal}
\end{eqnarray} 
Here, $k_{\rm B}$ is the Boltzmann constant and $T$ is the temperature. 
Thus, reaching the SQL needs the condition: 
\begin{eqnarray}
R_{\rm s}=\frac{S^{(2)}_{\rm FF,q}}{S^{(2)}_{\rm FF,th}}=|G_{\rm opt}|^2/(n_{\rm th}\kappa\gamma_{\rm m})>1,
\end{eqnarray}
where $n_{\rm th}$ is the phonon occupation number. 
To reduce the thermal noise, one can freely suspend a massive mirror in order to allow the mirror to be isolated from the environment. 
The pendulum motion of the suspended mirror is dominantly trapped by the gravitational potential, and thus the dissipation of the pendulum is gravitationally diluted by a factor of $k_{\rm grav}/k_{\rm el}=4l\sqrt{mg/\pi Y}/r^2$ (in this paper, we call it ``Q enhancement factor'') \cite{saulson1990}, where $k_{\rm grav}$ and $k_{\rm el}$ are the gravitational and elastic spring constants of the pendulum, $r$ is the radius of the wire, $l$ is the length of the wire, $m$ is the mass of the mirror, $Y$ is the Young's module of the wire, and $g$ is the gravitational acceleration. 
From Eq. (\ref{thermal}), any reduction of the dissipation results in a reduction of a thermal fluctuation force, which also drives the mechanical motion similarly to the quantum back-action, by a factor of $k_{\rm grav}/k_{\rm el}$. 

\begin{figure}[h]
\includegraphics[width=4in]{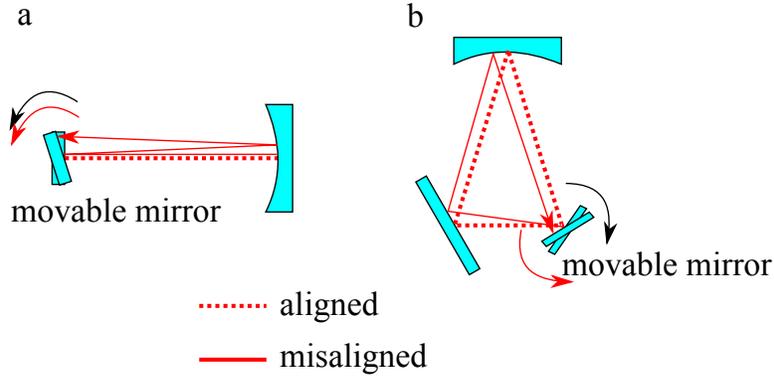}
\centering
\caption{{\bf Schematic of optical torsional effects.}
The schematic responses of the optical axes to the angular motion of the movable mirror are shown. 
The detailed response for the triangular cavity is given in Ref. \cite{kawazoe2011}. 
An optical torque occurs through the stationary radiation pressure. 
{\bf a:} In the case of the linear optical cavity, the optical torque occurs in the same direction as the angular motion. 
This results in an anti-restoring force. 
{\bf b:} In the case of the triangular optical cavity, the optical torque occurs in the opposite direction as the angular motion. 
This results in a restoring force.}
\label{fig1}
\end{figure}

Although this isolation largely reduces the suspension thermal noise, a stationary radiation force of the light exposes the free mass to instability through an optical torsional anti-spring effect as shown in Fig. \ref{fig1}(a), which is given by $\kappa_{\rm opt}=-P_{\rm circ}L_{\rm round}/c$, in conventional experiments utilizing a linear optical cavity \cite{sidles2006,sakata2010} (we suppose that the g-parameter closes to zero, which results in the smallest anti-restoring force). 
Here, $P_{\rm circ}$ is the intra-cavity power, $L_{\rm round}$ the round-trip length of the cavity, and c the speed of light. 
The stable condition concerning both the mechanical restoring force of the wire $\kappa_{\rm wire}$ and the optical anti-restoring force $\kappa_{\rm opt}$ is given by
\begin{eqnarray}
\kappa_{\rm wire}+\kappa_{\rm opt}>0. 
\label{condition}
\end{eqnarray}
Thus, $\pi G c/(2lL_{\rm round})>P_{\rm circ}/r^4$ should be satisfied in the case of using a single wire to suspend, where $G$ is the modulus of rigidity of the wire. 
This technical limitation becomes a significant issue, because a fundamental compromise between the stability and the sensitivity is generated; sufficient tolerance with firm suspension increases dissipation of the pendulum through the decrease of the gravitational dilution, which results in the increase of the thermal fluctuating force.

\begin{figure}[h]
 \begin{minipage}{0.5\hsize}
  \begin{center}
   \includegraphics[width=2.9in]{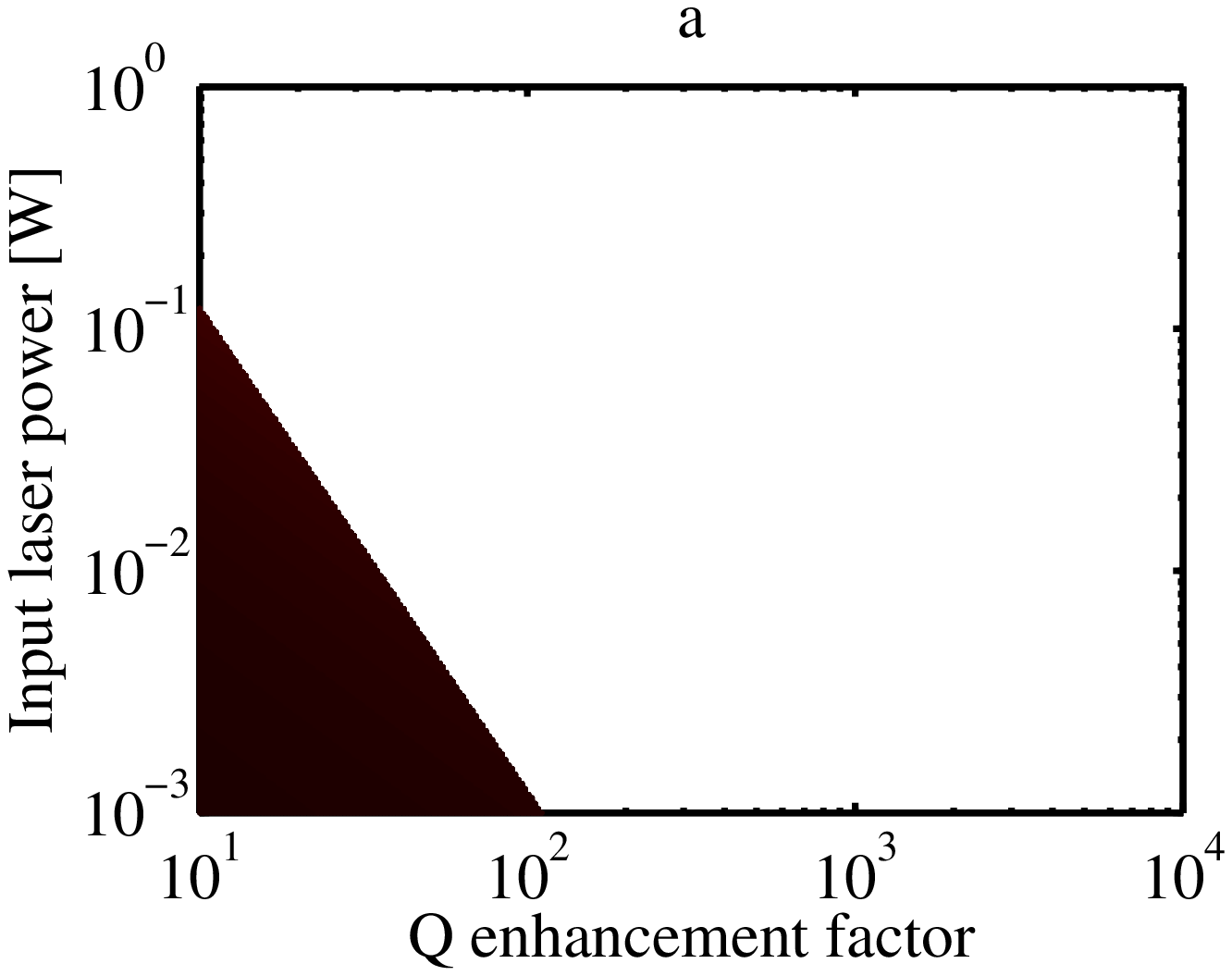}
  \end{center}
 \end{minipage}
 \begin{minipage}{0.5\hsize}
  \begin{center}
   \includegraphics[width=2.9in]{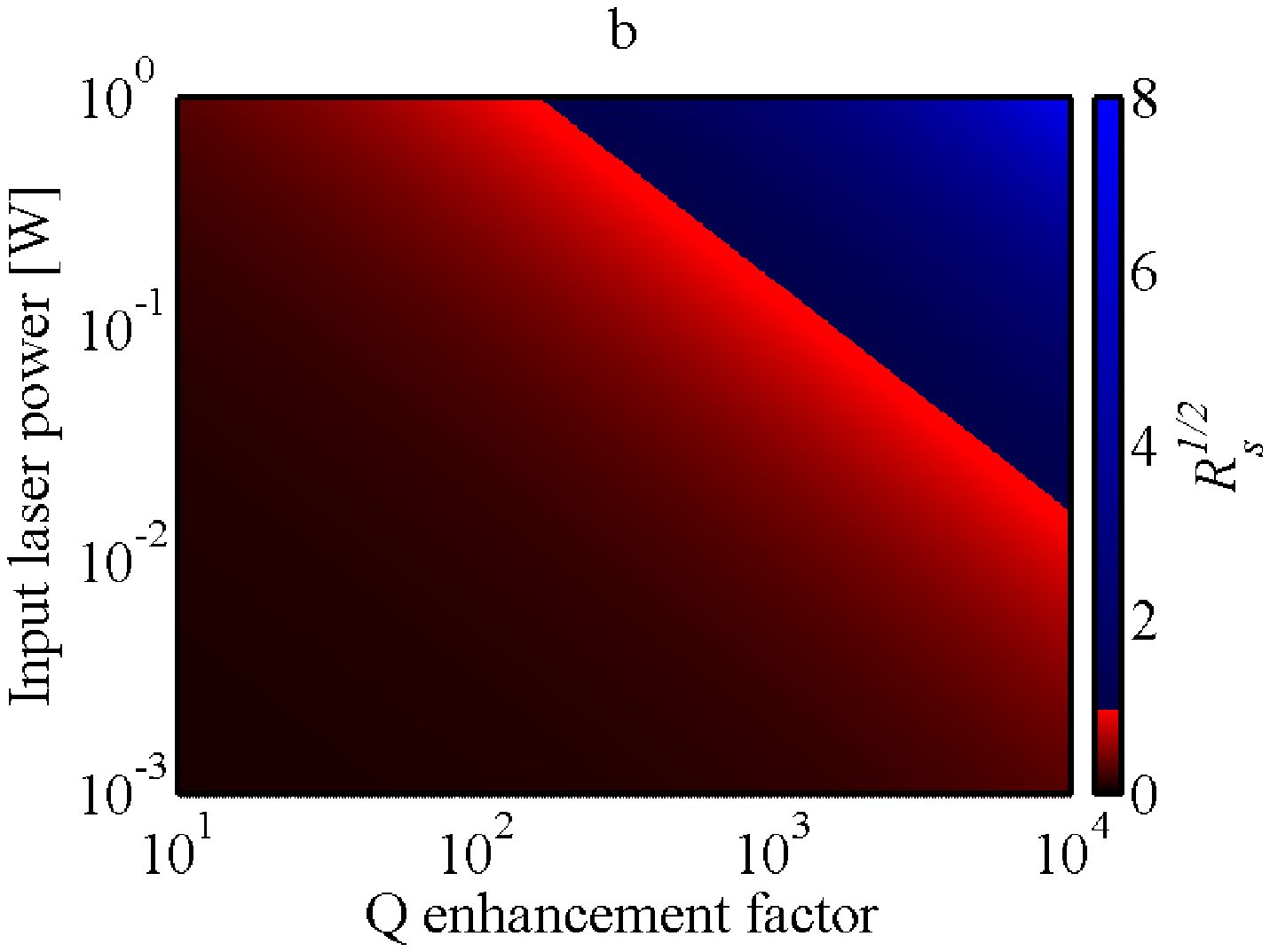}
  \end{center}
 \end{minipage}
 \caption{{\bf The square root of the ratio of the quantum back-action to the thermal fluctuating force.} 
{\bf a:} Trade-off relationship between the sensitivity and the stability is generated, in the case of using a linear cavity. 
A white domain represents the area, where $\kappa_{\rm wire}+\kappa_{\rm opt}<0$. 
Parameters: $l=0.02$ m, $m=14.7$ mg, $\sigma=0.28$, $T=300$ K, and $\kappa/(2\pi)=1.39$ MHz. 
Here, we suppose that each dissipation mechanisms are due to viscous friction (frequency dependent friction), for simplicity. 
In addition, we suppose that the intrinsic quality factor of the wire of $Q_{\rm in}$ is 3,800 (i.e., $Q_{\rm m}=Q_{\rm en}\times Q_{\rm in}$), and it has no dependence of the radius of the wire.
{\bf b:} As for a triangular cavity, the square root of the ratio is plotted. 
The ratio of $R_{\rm s}$ becomes one at the boundary of blue and red domains. 
In the blue domain, there is a possibility to reach the SQL. 
The same parameters as in Fig. \ref{fig2}(a) are used, but there is no unstable domain. 
In addition, the incident angle of $\beta$ is supposed to be 0.64 rad.
}
\label{fig2}
\end{figure}

As a result, the enhancement of $R_{\rm s}$ due to the gravitational dilution is limited by
\begin{eqnarray}
Q_{\rm en}<\sqrt{\frac{G}{Y}8lmg\frac{1}{\tau P_{\rm circ}}}=\sqrt{\frac{2lmg}{(1+\sigma)}\frac{\kappa}{P_{\rm in}}}.
\label{limitation}
\end{eqnarray}
Here, $Q_{\rm en}$ is the Q enhancement factor, and $\tau$ the round-trip time, $\sigma$ the Poisson's ratio, and $P_{\rm in}$ the input laser power. 
The ratio of $R_{\rm s}$ in the case of using the linear cavity is plotted as a function both of the input laser power and the Q enhancement factor in Fig. \ref{fig2}(a). 

In this paper, we describe using a triangular optical cavity to overcome this limitation, and then propose an experiment to reach the SQL. 

\section{Optical torsional spring effect in the triangular cavity}
In order to overcome the trade-off given by Eq. (\ref{limitation}), Eq. (\ref{condition}) should be voluntary satisfied. 
Then, we focus on the geometrical difference between a linear and a triangular cavity. 
Unlike the linear cavity, light experiences odd numbers of reflections on mirrors inside the triangular cavity, as shown in Fig. \ref{fig1}(b).
This results in a {\it positive} torsional spring effect. 
Figure \ref{fig1} enables us to intuitively and visually understand the difference between the linear and triangular optical cavities. 
The positive torsional spring effect overcomes the trade-off relationship written by Eq. (\ref{limitation}), as shown in Fig. \ref{fig2}(b). 
Although one can calculate the optical positive torsional effect using the result described in Ref. \cite{kawazoe2011}, we use the result described in Ref. \cite{sigg2003} for simplicity. 

\section{Model of a triangular cavity}
\begin{figure}[h]
\centering
\includegraphics[width=1in]{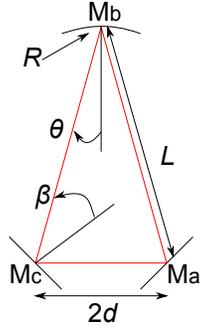}
\caption{{\bf Schematic of the triangular cavity.}
This figure represents the layout of the triangular cavity. 
The triangular cavity formed by two flat mirrors, labeled ${\rm M_{\rm a}}$ (the movable mirror) and ${\rm M_{\rm c}}$, and a curved mirror, labeled ${\rm M_{\rm b}}$. 
$L$, represents the distance between the curved mirror and the flat mirror; $d$ is half the distance between the two flat mirrors, $R$ is the radius of curvature of mirror ${\rm M_{\rm b}}$, $\theta$ is the incident angle on the curved mirror, and $\beta$ is the incident angle on the flat mirror.}
\label{fig3}
\end{figure}
In this section, we derive the torsional spring effect around the suspension axis (yaw) due to the radiation-pressure torque in a triangular cavity. 
We begin by considering a two-dimensional triangular cavity formed by two flat mirrors, labeled ${\rm M_{\rm a}}$ and ${\rm M_{c}}$, and a curved mirror, labeled ${\rm M_{\rm b}}$, as shown in Fig. \ref{fig3}.
We decompose the rotations of the two flat mirrors into two basis modes: the common-mode (same the rotation direction, the same amount) and the differential-mode (opposite rotation direction, the same amount).
Any misalignment state of the two mirrors can be expressed as a linear combination of these two basis modes.
In this picture, the relationship between the misalignment angle, $\Delta \alpha$, of the basis modes and the change in beam position on each of the mirror, $\Delta x$, is given by \cite{sigg2003}
\begin{eqnarray}
	{\Delta{\boldmath {x}}}=LK_h\Delta{\bf{\alpha}}
\end{eqnarray}
with
\begin{eqnarray}
K_h=\frac{1}{L(d+L-R)}
\left(
\begin{array}{ccc}
\frac{-2d(L-R)}{\cos\beta} & 0 & \frac{-dR}{\cos\beta} \\
0 & \frac{-2L(d+L-R)}{\cos\beta} & 0 \\
-dR & 0 & (d+L)R \\
\end{array} 
\right)
\end{eqnarray}
Here, $L$ is the distance between the curved mirror and the flat mirror, $d$ is half the distance between two flat mirrors, $R$ is the radius of curvature of the mirror ${\rm M_b}$, and $\beta$ is the incident angle on the flat mirror.  
The torque, $N_{\rm rad}$, on each mirror induced by the radiation pressure is given by
\begin{eqnarray}
N_{\rm rad}=\frac{2P_{\rm circ}}{c}LTK_h
\end{eqnarray}
with
\begin{eqnarray}
T=
\left(
\begin{array}{ccc}
\cos\beta & 0 & 0 \\
0 & \cos\beta & 0 \\
0 & 0 & \cos\theta \\
\end{array} 
\right)
\label{torque}
\end{eqnarray}
where $\theta$ is the incident angle on the curved mirror, and $\beta$ is the incident angle on the flat mirror.

For simplicity, we consider the situation where only the mirror ${\rm M_{\rm a}}$ is movable and others are fixed. 
In this case, the equations of motion are given by
\begin{eqnarray}
I_{\rm a}\ddot{\alpha}_{\rm a}=-(\kappa_{\rm opt}+\kappa_{\rm wire})\alpha_{\rm a}, \label{model}
\end{eqnarray}
\begin{eqnarray}
\kappa_{\rm opt}=-\frac{2P_{\rm circ}L}{c} \frac{(d+L)[R-(L+dL/(d+L))]}{L(d+L-R)},
\label{reso}
\end{eqnarray}
where, $I_{\rm a}$ is the moment of inertia about the wire axis of mirror ${\rm M_{\rm a}}$, $\kappa_{\rm opt}$ is the angular spring constant of mirror ${\rm M_{\rm a}}$ induced by the radiation pressure, and $\kappa_{\rm wire}$ is the mechanical torsional spring constant of mirror ${\rm M_{\rm a}}$ in yaw. 
Under the self-consistent condition of the cavity, which is given by $0<d+L<R\cos(\theta)$, Eq. (\ref{reso}) is always positive. 
Thus, this configuration has intrinsic stability in the yaw direction. 

From this equation, we can derive the resonant frequency of the yaw motion as
\begin{eqnarray}
f_{\rm a}=\frac{1}{2\pi}\sqrt{\frac{\kappa_{\rm opt}+\kappa_{\rm wire}}{I_{\rm a}}}\label{theory}
\end{eqnarray}
From Eqs. (\ref{reso}) and (\ref{theory}), it is found that the angular resonant frequency is increased with  increased circulating power.

\section{Experiment}
\begin{figure}[h]
\centering
\includegraphics[width=5in]{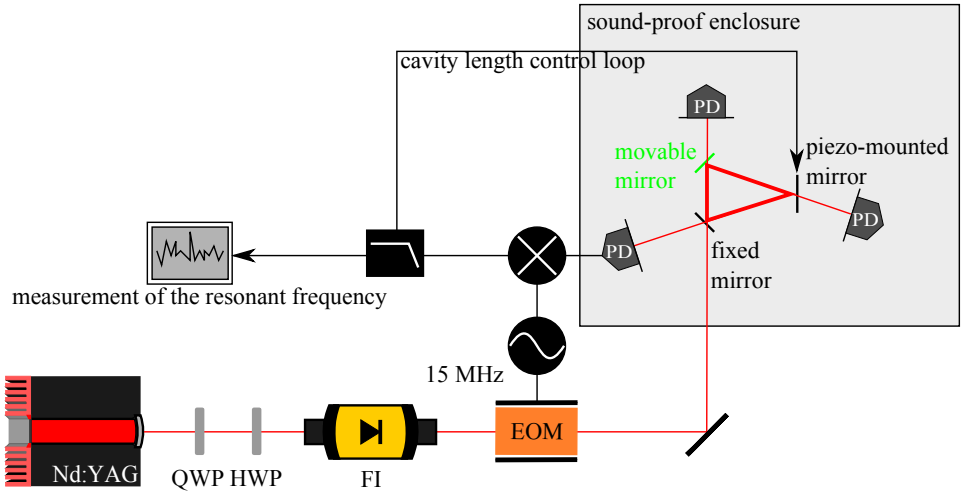}
\caption{{\bf The detailed experimental setup for observing optical torsional spring effect.}
The laser beam (red line) was fed into the triangular cavity.  
An electro-optical modulator (EOM) was used to apply frequency sidebands for a Pound-Drever-Hall (PDH) method \cite{drever1983}.
Light was detected at various points using photodetectors (PD).
HWP, Half-Wave Plate; QWP, Quarter-Wave Plate; FI, Faraday Isolator.}
\label{fig4}
\end{figure}
In order to quantitatively verify the model described in the previous section, we measured the angular resonant frequency of a mirror in a triangular cavity.
 By changing the internal power of the cavity, thus changing $\kappa_{\rm opt}$, we expect the resonant frequency to change according to Eq. (\ref{theory}). 
 
A schematic of the experimental setup is shown in Fig. \ref{fig4}.
The laser source was a monolithic non-planar Nd:YAG ring laser with a 2 W continuous-wave single-mode output power at 1064 nm.
We used an electro-optic modulator (EOM) as a phase modulator at 15 MHz to lock the triangular cavity  via a Pound-Drever-Hall (PDH) locking scheme \cite{drever1983}.
The triangular cavity with a length of 100 mm and a finesse of 223 ($\kappa_{\rm in}/\kappa=0.48$, i.e., intra-cavity power gain is 69, where $\kappa_{\rm in}$ is the decay rate for the input coupler.) was composed of two flat mirrors and a fixed curved mirror with a radius of curvature of 75 mm.
One of the two flat mirrors was a half-inch fused silica mirror suspended by a tungsten wire of 20 $\mu$m diameter and 40 mm length.
The suspended mirror was attached to an oxigen-free copper cylinder of 3 mm diameter and 3 mm  thickness, which was damped by an eddy-current using a doughnut-shaped magnet.
Because of its shape, the magnet damps only the pendulum motion without decreasing the mechanical quality factor of the yaw motion. 
The resonant frequency of the yaw motion was measured to be 369 mHz, by optical shadow sensing. 
The curved mirror was fixed, and was mounted on a piezoelectric transducer (PZT; NEC Tokin, AE0505D08F), which was used as an actuator to keep the cavity in resonance with the laser.
The triangular cavity and photodetectors (HAMAMATSU, G10899-01K, InGaAs photodiode) were placed in a vacuum desiccator (AS ONE, 1-070-01) for acoustic shielding. 

The reflected light was received by a photo-detector, and its output signal was demodulated at the modulation frequency.
This signal was then low-pass filtered with a cutoff frequency of 1 Hz, and fed back to the PZT actuator.
The unity gain frequency of the length control servo was approximately 1 kHz.
We used this signal to stabilize the cavity length and also to measure the angular (yaw) resonant frequency.
The yaw motion of the suspended mirror generated the PDH signal, because there was a slight miscentering of the beam position on the suspended mirror.
The transmitted light was also detected in order to measure the finesse of the triangular cavity. 
The incident light power into the cavity was varied from 60 mW to 1 W in order to measure the change in the  angular resonant frequency.

\section{Results}
\begin{figure}[h]
 \begin{minipage}{0.5\hsize}
  \begin{center}
   \includegraphics[width=2.8in]{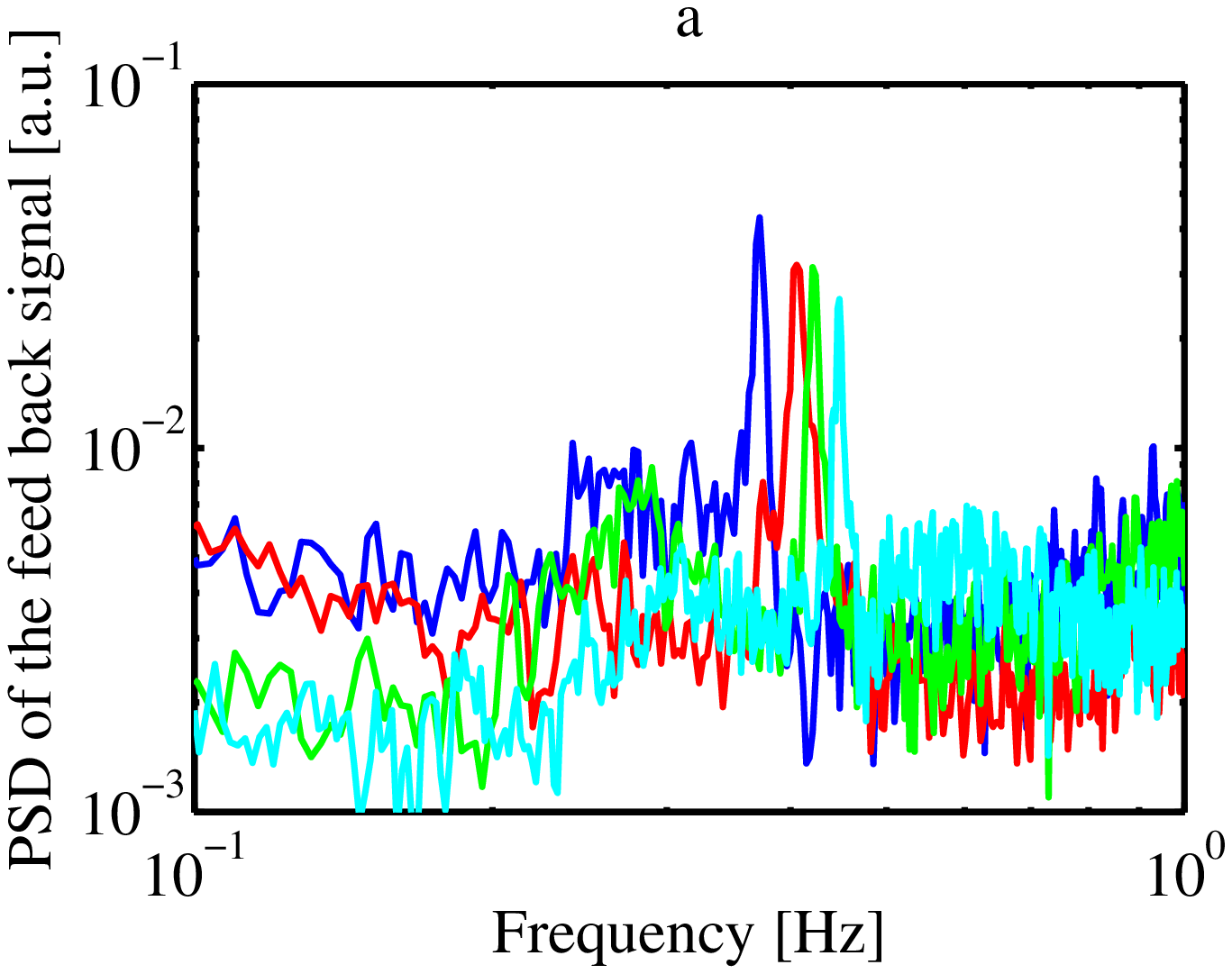}
  \end{center}
  \label{fig5_1}
 \end{minipage}
 \begin{minipage}{0.5\hsize}
  \begin{center}
   \includegraphics[width=2.8in]{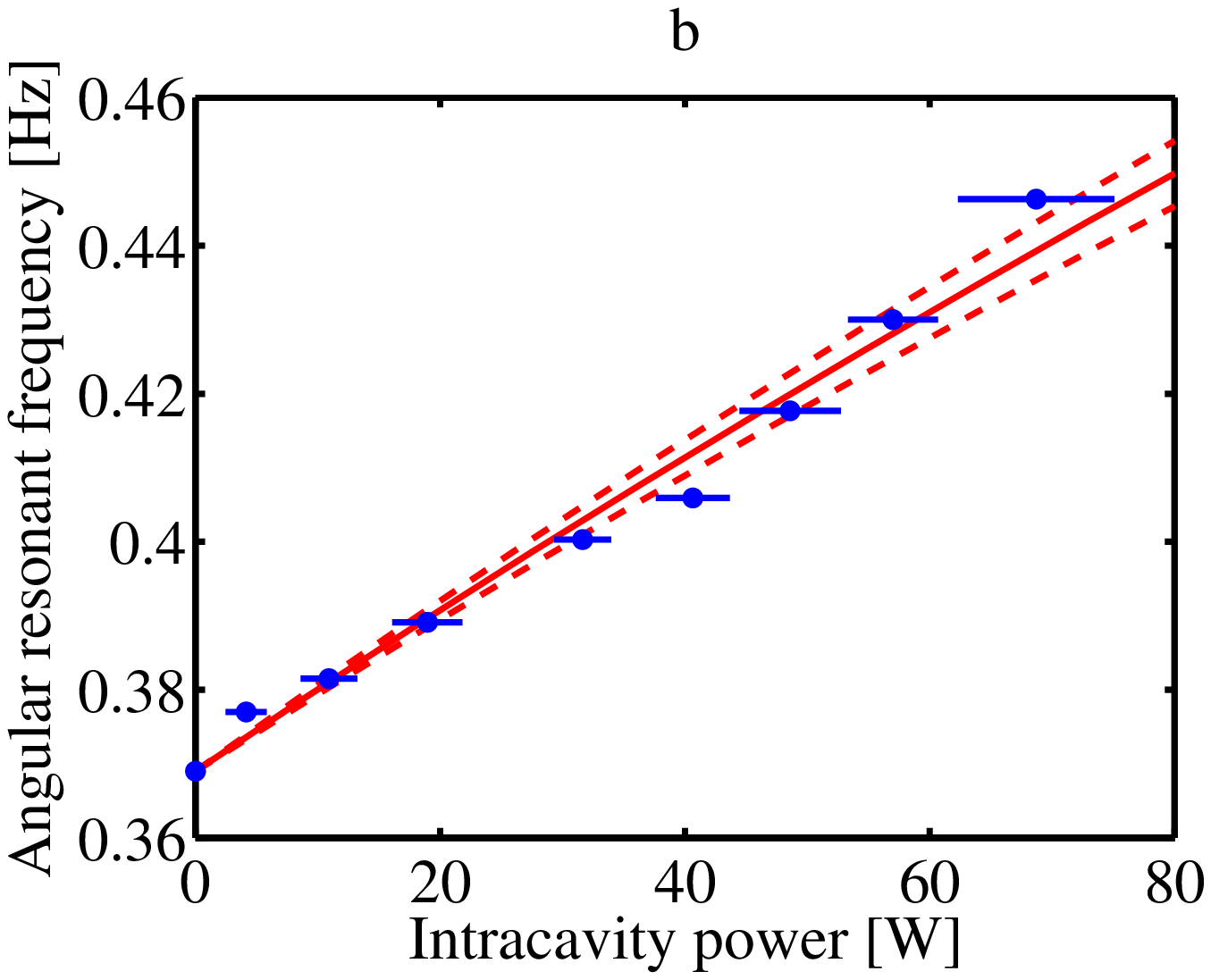}
  \end{center}
  \label{fig5_2}
 \end{minipage}
 \caption{{\bf Measurement of the optical torsional spring in the triangular cavity.} 
{\bf a:} The observed power spectral density (PSD) of the feedback signal. 
A peak represents the yaw motions of the suspended mirror at the intra-cavity powers [4 W (blue), 32 W (red), 46 W (green), and 69 W (cyan)].
The yaw resonant frequency is shifted to the higher one with the increased power.
{\bf b:} Angular resonant frequency of the mirror suspension against the intra-cavity power.
The blue circles are the measurement data and the blue horizontal lines are statistical errors.
The red curve is the theoretical one from Eq. (\ref{theory}), and the dashed red curve shows systematic error.}
\end{figure}
Figure \ref{fig5_1} (a) shows the observed spectra of the feedback signal with the intra-cavity power at 4 W (blue), 32 W (red), 46 W (green), and 69 W (cyan).
The peaks at around 0.4 Hz are the yaw resonances.
The angular resonant frequency increases with increasing circulating power.
The measured angular resonant frequencies are plotted against the intra-cavity power in Fig. \ref{fig5_2}(b).
The blue circles are the measured values and the horizontal lines are the statistical errors, which arises from the uncertainty of $l$ and $L$. 
The dashed red curves are the theoretical predictions, obtained from Eqs. (\ref{reso}) and (\ref{theory}) with $L=(4.4\pm0.1)\times10\ {\rm mm}$, $d=(1.0\pm0.1)\times10\ {\rm mm}$, $\beta=0.7\pm0.1\ {\rm rad}$, and $\kappa_{\rm opt}=(3.9\pm0.2)\times10^{-10}\times P_{\rm circ}\ {\rm Nm/rad}$. 
The theoretically calculated values show good agreement with the experimental results, which suggests that Eq. (\ref{reso}) is suitable for modeling the torsional spring effect caused by the optical restoring force.

\section{Discussions}
From this result and the following consideration, the advantage of the triangular cavity can be understood. 
When the linear cavity is used, the instability can be mitigated by, e.g.,: (i) reducing the optical power; (ii) shortening the cavity length; (iii) using a thick wire for suspension; (iv) using multiple wires for suspension; (v) using the long wire; (vi) active control; and (vii) using a linear optical cavity that consists of fixed and suspended mirrors under the {\it negative}-g condition (i.e., both focal points are inside the cavity; in other words, both mirrors have a concaved structure). 
However, those induce: (i) a reduction of the quantum back-action;  (ii) an increase of the linewidth of the cavity (i.e., reduction of laser frequency noise); however, in practice it is insufficient only by it; (iii) a reduction of the gravitational dilution (i.e., increasing the suspension thermal noise); (iv) introducing an unexpected thermal noise through the unexpected normal mode generated by the complicated suspension system \cite{neben2012}; (v) a decrease of the resonant frequency of the violin mode;  (vi) the necessity of using a more macroscopic mirror to be attached along with the actuator, which would result in decreasing the SQL (i.e., relatively increase of all technical noise), and also might introduce some other dissipation through the actuator (to avoid these issues, using the lossless control system via radiation pressure without attached actuators has been proposed \cite{kawamura}); and (vii) the necessary of using a sufficiently concaved and small mirror in order to make the cavity length shorten and avoid the same issue of the first part of (vi). 
(If an appropriate mirror can be manufactured, this method has no issue. We have been trying, but it is still challenging.)
On the other hand, the triangular cavity has an {\it intrinsic} stability in yaw direction. 
As a result, one can conclude that the triangular cavity overcomes the fundamental compromise. 

Until now, we have only paid attention to the yaw.
Note that this is a sufficient discussion in order to consider the stability of our triangular cavity.
Because the suspended mirror can easily have sufficient mechanical positive torsional spring constants for a pitch  without increasing the suspension thermal noise, even though the anti-torsional spring effect occurs for the pitch.
This is due to the fact that the stiffness of the pitch does not depend on the radius of the wire, which mainly determines the Q enhancement factor, but depends on the radius of the mirror. 
Also, as for damping effect, it is generally stable, because phase delay of the optical torque is generally much smaller than that of the mechanical torque in the case of using a short length cavity, and thus the optomechanical damping effect can be negligible. 

\section{Toward reaching the SQL}
\begin{figure}[h]
\centering
\includegraphics[width=3.9in]{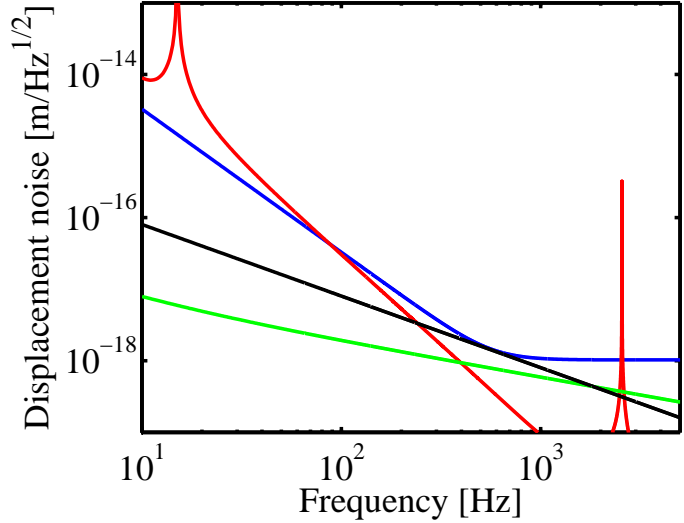}
\caption{{\bf Design sensitivity for reaching the SQL.}
The quantum noise (blue), the suspension thermal noise (red), the mirror thermal noise (green), and the SQL (black)  are shown. 
The same parameters as in Fig. \ref{fig2} are used. 
The peak at around 10 Hz is the rocking mode and at around 2 kHz is the 1-st violin mode of the wire, respectively. 
}
\label{fig6}
\end{figure}
Our demonstration shows that the triangular cavity is intrinsically stable in yaw direction, and thus it can become an effective system to reach the SQL. 
Figure \ref{fig6} shows a design sensitivity for reaching the SQL, by using the triangular cavity composed of one movable mirror and two fixed mirrors.
We consider a tiny mirror of 5 mm in diameter and 0.3 mm in thickness, leading to a mass of 14.7\ mg.
The tiny mirror is suspended by a single thin tungsten wire (1 $\mu$m in diameter and 20 mm in length).
Here, we suppose the dissipation mechanism is due to internal friction (frequency independent internal friction). 
In addition, we suppose that the intrinsic quality factor of the wire is 3,800, since the mechanical quality factor of the tungsten wire with 3 $\mu$m in diameter was measured to be 3,800 \cite{matsumoto2013}.
The quality factor of pendulum can be improved up to $2\times10^7$ due to the gravitational dilution.
In this case, suspension thermal noise is shown in Fig.\ref{fig6} as a red line.
Mechanical quality factors of the coating and the silica structure of the movable mirror are estimated to be $1\times10^4$ and $1\times10^6$, respectively. 
The mirror thermal noise (sum of the coating and the substrate thermal noise) is shown in Fig.\ref{fig6} as a green line.
The blue line shows the quantum noise (i.e., the quantum back-action and the shot noise). 
The input laser power is 5 mW, which results in the intra-cavity power of 3.2 W.
Although the laser power is about 10,000-times higher than the instability limit for the linear cavity, it is relatively weak. 
This is because that the frequency, where the SQL equals the sum of the quantum noises ($f_{\rm eq}\simeq2G_{\rm opt}\sqrt{\hbar/(\kappa m)}/(2\pi)\simeq 600$ Hz) should be smaller than the frequency, where the SQL equals the mirror thermal noise ($\simeq$ 2 kHz). 


As a result, the SQL can be reached in the range of about 240 to 1,700 Hz, even at the room temperature. 
\section{Conclusion}
We measured a radiation-pressure induced angular (yaw) spring effect in the triangular optical cavity.
Theoretical calculation shows good agreement with the experimental results.
This result suggests that the triangular cavity is intrinsically stable in yaw direction, and thus it is possible to store high power beam in the triangular cavity with a large gravitational dilution. 
Therefore, the triangular cavity can overcome the trade-off between the stability and the sensitivity concerning the linear cavity. 
Moreover, we propose the experiment using the triangular cavity, in order to reach the standard quantum limit at beyond the resonant frequency of the pendulum.

\section*{Acknowledgments}
We are particularly grateful to S. Kawamura, Y. Shikano, N. Ohmae and also to W. Kokuyama for stimulating and informative discussions.
This research was supported by a Grant-in-Aid for JSPS Fellows No.
$25\cdot10490$.
\end{document}